\documentclass[pra,nofootinbib,amssymb,amsmath]{revtex4}
\usepackage[colorlinks,citecolor=blue,linkcolor=blue]{hyperref}

\usepackage{appendix}
\usepackage{graphicx}
\usepackage{amssymb,amsmath,amsbsy} 
\usepackage{color}
\usepackage[utf8]{inputenc}
\usepackage[polish,english]{babel}
\usepackage[T1]{fontenc}

\def\B#1{\!\left(#1\right)}
\def\BB#1{\!\left[#1\right]}

\def\O#1{O\left(J^{#1}\right)}

\def\be{\begin{equation}}
\def\ee{\end{equation}}

\def\bx{\begin{equation}\begin{aligned}}
\def\ex{\end{aligned}\end{equation}}

\def\bee{\begin{equation*}}
\def\eee{\end{equation*}}

\def\for{\ \ {\rm for} \  }

\def\bg{\begin{equation}\begin{gathered}}
\def\eg{\end{gathered}\end{equation}}

\def\bgg{\begin{equation*}\begin{gathered}}
\def\egg{\end{gathered}\end{equation*}}

\def\la{\langle}
\def\ra{\rangle}

\begin{document}

\title{Properties of the one-dimensional Bose-Hubbard model from a high-order perturbative expansion}

\author{Bogdan Damski and Jakub Zakrzewski}
\affiliation{Instytut Fizyki imienia Mariana Smoluchowskiego, Uniwersytet Jagiello{\'n}ski, ulica  {\L}ojasiewicza 11, 30-348 Krak\'ow, Poland}

\begin{abstract}
We employ a  high-order perturbative expansion  to characterize the ground state of the
Mott phase of the one-dimensional Bose-Hubbard model. We compute  for different integer 
filling factors 
the energy per lattice site,
the two-point and density-density correlations, and expectation values of powers
of the on-site number operator determining the local atom  number fluctuations (variance, skewness, kurtosis). 
We compare these expansions to numerical
simulations of the infinite-size system to determine their range of
applicability. We  also discuss a new sum rule for the density-density  correlations that can be used in  both 
equilibrium and non-equilibrium systems. 
\end{abstract}
\maketitle

\section{Introduction}
The Bose-Hubbard models capture  key properties  of numerous  experimentally-relevant configurations 
of  cold bosonic atoms placed in  optical lattices 
\cite{JakschAnnPhys,LewensteinAdv,KubaReview,KrutitskyReview}. The simplest 
of them is defined by the Hamiltonian
\bg
\label{H}
\hat{H}= -J \sum_{i}\B{\hat{a}_{i+1}^\dag \hat{a}_i + {\rm h. c.}}
+\frac{1}{2} \sum_{i}\hat{n}_i\B{\hat{n}_i-1}, \\
[\hat{a}_i,\hat{a}_j^\dag]=\delta_{ij}, \ [\hat{a}_i,\hat{a}_j]=0, 
\  \hat n_i=\hat{a}_i^\dag\hat{a}_i,
\eg
where the first term describes  tunnelling  between
adjacent sites, while the second one accounts for  on-site interactions. The
competition between these two terms leads to the  Mott
insulator--superfluid  quantum phase transition when the filling factor
(the mean number of atoms per lattice site) is integer \cite{Fisher89,Sachdev}. The system is in the
superfluid phase when the tunnelling term dominates ($J>J_c$) whereas
it is in the Mott insulator phase when the interaction term wins out
($J<J_c$). The location of the critical point depends on the filling factor
$n$ and the dimensionality of the system. We consider the one-dimensional
model (\ref{H}), where it was estimated that 
\be
J_c\approx\left\{
\begin{array}{ll}
0.3 & \for n=1 \\
0.18 & \for n=2 \\
0.12  & \for n=3
\end{array}
\right..
\label{Jc}
\ee
It should be mentioned  that there is a few percent disagreement
between different numerical computations of the position of the critical 
point (see Sec. 8.1 of Ref. \cite{KrutitskyReview} for an exhaustive discussion of 
this topic). That  affects neither our results nor  the discussion of our
findings.

The Bose-Hubbard model (\ref{H}), unlike some one-dimensional spin and cold atom  systems \cite{Bethe,Sachdev}, 
is not exactly solvable. Therefore, it is not surprising that accurate  analytical results 
describing its properties  are scarce.  To the
best of our knowledge, the only systematic way of obtaining them is 
provided by the perturbative expansions 
\cite{FreericksPRB1996,MonienPRB1999,BDJZPRA2006,EckardtPRB2009,AEckardtPRB2009,HolthausEPL2015,TrivediPRA2009}. 
In addition to delivering  (free of finite-size effects) insights into physics of the
Bose-Hubbard model, these expansions can be used to benchmark approximate
approaches (see e.g. Refs. \cite{KnapPRB2010,KnapPRA2012}).

We compute the following ground-state expectation values: 
the energy per lattice site $E$, the two-point correlations
$C(r)=\langle\hat{a}_j^\dag\hat{a}_{j+r}\rangle$, the density-density correlations
$D(r)=\langle \hat{n}_j\hat{n}_{j+r}\rangle$, and the powers  of the on-site number
operator $Q(r)=\la\hat{n}^r_i\ra-n^r$. 

Our perturbative expansions are obtained with the technique described in Ref. \cite{BDJZPRA2006} 
(see also Ref. \cite{EckardtPRB2009} for a similar approach yielding the same results).
The   differences with respect to Ref. \cite{BDJZPRA2006} are the
following. First, we have computed   perturbative expansions for the filling factors $n=2$ and $3$, 
which were not studied   in Ref. \cite{BDJZPRA2006}. Second, we have enlarged the
order of all the  expansions for the $n=1$ filling factor that were reported earlier. Moreover,
several perturbative results for the $n=1$ case, that were not listed in Ref.
\cite{BDJZPRA2006},  are provided in Appendix \ref{sec_appone}. 
Third, we have computed perturbative expansions for the expectation values 
of different powers of the on-site atom number operator, which were not discussed in Ref. \cite{BDJZPRA2006}. 
This allowed us for computation of the skewness and kurtosis characterizing 
on-site atom number distribution. Fourth, we have derived an important  sum rule for the density-density 
correlations allowing for  verification of all our perturbative
expansions  for these correlations.

The range of validity of our perturbative expansions is carefully established
through numerical simulations. There is another crucial difference here with respect to our 
former work \cite{BDJZPRA2006}. Namely, instead of considering a $40$-site
system, we study an infinite system using the translationally invariant version of the Time Evolving 
Block Decimation (TEBD) algorithm  sometimes referred to as iTEBD \cite{Vidal07} (where i stands for infinite).
The ground state of the system is found by imaginary time propagation
\cite{KubaDom}.
For the detailed description of the method and its relation to the 
density matrix renormalization group studies see 
the excellent review \cite{Schollwoeck11}. 
The application of iTEBD
allows for obtaining results free of the finite-size effects
from numerical computations (see Appendix \ref{sec_appitebd} for the details
of these simulations).
Our symbolic perturbative expansions have been done on a $256$ Gb computer.
The numerical computations require two orders of magnitude smaller computer memory.

The outline of this paper is the following. We discuss in Sec. \ref{sec_id} 
various identities that can be used to check the validity of our
perturbative expansions. In particular, we derive there a sum rule for
density-density correlation functions. Sec. \ref{sec_energy} is focused on  the
 ground state energy per lattice site. Sec. \ref{sec_variance} shows our results
for the variance of the on-site atom number operator. 
Sec. \ref{sec_powers} discusses expectation value of different 
powers of the on-site number operator and the related observables: the skewness and kurtosis
of the local atom number distribution.
Sec. \ref{sec_twopoint} discusses 
the two-point correlation functions. 
Sec. \ref{sec_density} provides results on the density-density correlations. 
The perturbative expansions presented in Secs.
\ref{sec_energy}--\ref{sec_density} are  compared to numerics, which allows
for establishing the range of their applicability. Additional perturbative expansions 
are listed in Appendices
\ref{sec_appone}, \ref{sec_apptwo}, \ref{sec_appthree} for the filling factors $n=1,2,3$,
respectively.  The paper ends with a brief summary (Sec. \ref{sec_sum}).

\section{Ground state  identities and sum rule}
\label{sec_id}
There are several identities rigorously verifying our perturbative results. First, straight from the 
eigen-equation  one  gets that the ground state energy per lattice site, $E$, satisfies
\be
 E= -2JC(1) + \frac{D(0)-n}{2}.
\label{ECD}
\ee
 It is easy to check that our perturbative
expansions
-- (\ref{E_n=1}), (\ref{var_n=1}), and (\ref{C_n=1_r=1}) for $n=1$;
(\ref{E_n=2}), (\ref{var_n=2}), and (\ref{C_n=2_r=1}) for $n=2$; and 
(\ref{E_n=3}), (\ref{var_n=3}), and (\ref{C_n=3_r=1}) for $n=3$ --
satisfy this identity.

Combining this result with the Feynman-Hellmann theorem, 
\bee
\frac{d}{dJ}E = \left\langle \frac{d\hat H}{dJ}\right\rangle,
\eee
we get 
\bee
\frac{d}{dJ}D(0) = 4J\frac{d}{dJ}C(1).
\eee
A similar identity can be found in Sec. 7.1 of Ref. \cite{KrutitskyReview}.
Once again, it is straightforward to check that  our expansions for $n=1,2,3$ 
satisfy this identity.

Finally, we obtain a sum rule for the density-density correlations in  a one-dimensional system
\be
D(0)-n^2+2\sum_{r=1}^\infty \BB{D(r)-n^2}=0.
\label{den-den}
\ee
It is again an easy exercise to check that our expansions -- 
(\ref{D_n=1_r=1})--(\ref{D_n=1_r=3}) and (\ref{D_n=1_r=4})--(\ref{D_n=1_r=8}) for $n=1$; 
(\ref{D_n=2_r=1})--(\ref{D_n=2_r=3}) and (\ref{D_n=2_r=4})--(\ref{D_n=2_r=6}) for $n=2$; 
and (\ref{D_n=3_r=1})--(\ref{D_n=3_r=3}) and (\ref{D_n=3_r=4})--(\ref{D_n=3_r=6}) for $n=3$
-- satisfy this sum rule \footnote{There is no need to perform the sum over infinite number of 
$D(r)$'s to see that our results satisfy the sum rule (\ref{den-den}). 
This follows from the observation that $D(r>0)-n^2=\O{2r}$. Thus, if our expansions for $n=1$
($n=2$ and $3$) are done up to the order $J^{16}$ ($J^{12}$), we need to know  
$D(r)$ only for $r=1,2,\cdots,8$ ($r=1,2,\cdots,6)$.}. Eq. (\ref{den-den}) can be obtained 
from the sum rule for the zeroth moment of the dynamic structure factor (see Ref.
\cite{PitaevskiiStringari} for  a general introduction to a dynamic structure
factor and its sum rules  and Ref.
\cite{RothJPB2004} for their discussion in  a Bose-Hubbard model). We have, however,
derived it in the following elementary way.

Consider a system
of $N$ atoms placed in the $M$-site periodic lattice ($N,M<\infty$). 
Assuming
that the system   is prepared in an eigenstate of the number operator, say $|\Psi\ra$,
we have 
\be
N^2=\left\la\Psi\left|\B{\sum_{i=1}^M\hat n_i}^2 \right|\Psi\right\ra=
\sum_{i,j=1}^M\la\Psi|\hat n_i\hat n_j|\Psi\ra.
\label{qwerty}
\ee
The next step is to assume that the correlations $\la\Psi|\hat n_i\hat n_j|\Psi\ra$ depend only on the 
 distance between the two lattice sites. This assumption 
allows for rewriting  Eq. (\ref{qwerty}) to the form 
\be
D(0)-\B{\frac{N}{M}}^2+2 \sideset{}{'} \sum_{r=1}^{\lfloor M/2\rfloor}\BB{D(r)-\B{\frac{N}{M}}^2}=0,
\label{finite-den}
\ee
where $\lfloor x\rfloor$ stands for the  largest integer not greater than $x$,
$\lfloor M/2\rfloor$ is the largest distance between  two lattice sites  in the $M$-site  periodic lattice,
and the prime in the sum indicates that in even-sized systems the summand for 
$r=\lfloor M/2\rfloor$  has to be multiplied by a factor $1/2$. 
One obtains Eq. (\ref{den-den}) by taking the limit of $N,M\to\infty$ such that the 
filling factor $n=N/M$ is kept constant. Such a procedure is
meaningful as long as the correlations $D(r)$ tend to $n^2$ sufficiently fast as
$r$ increases,
which we assume. 
The extension of the above sum rule to two- and three-dimensional systems is
straightforward, so we do not discuss it.

Instead, we mention that  the sum rule (\ref{finite-den}) can be also applied 
to non-equilibrium systems satisfying  the  assumptions used in its
derivation. It  can be  used either to study constraints on the 
dynamics of the density-density correlations 
or to verify the accuracy of 
numerical computations. Both applications are relevant for 
the studies of quench dynamics of the Bose-Hubbard model 
triggered by the  time-variation of the 
tunnelling coupling $J$ \cite{KollathJSTAT2008,KollathPRA2012,LaurantPRA2014}. 
We mention in passing that a completely different work on the sum rules applicable to the 
 Bose-Hubbard model can be found in Ref. \cite{KnapPRA2013}.

Finally, we mention that it has been shown in Ref. \cite{BDJZPRA2006} that
the ground state energy per lattice site and the density-density correlations in the Bose-Hubbard model 
are unchanged by the
\be
J\to-J
\label{JJ}
\ee
transformation, while the two-point correlations transform under (\ref{JJ}) as
$C(r)\to(-1)^rC(r)$. Using the same reasoning one can show that $Q(r)$ is symmetric
with respect to  (\ref{JJ}) as well. One can immediately check that all the
expansions that we provide satisfy these rules. This observation provides one
more consistency check of our perturbative expansions. Moreover, it  allows us to 
skip the $O(J^{m+2})$ term by the end of every expansion
ending with a $J^m$ term.

\section{Ground state energy}
\label{sec_energy}

\begin{figure}[t]
\includegraphics[width=0.56\textwidth,clip=true]{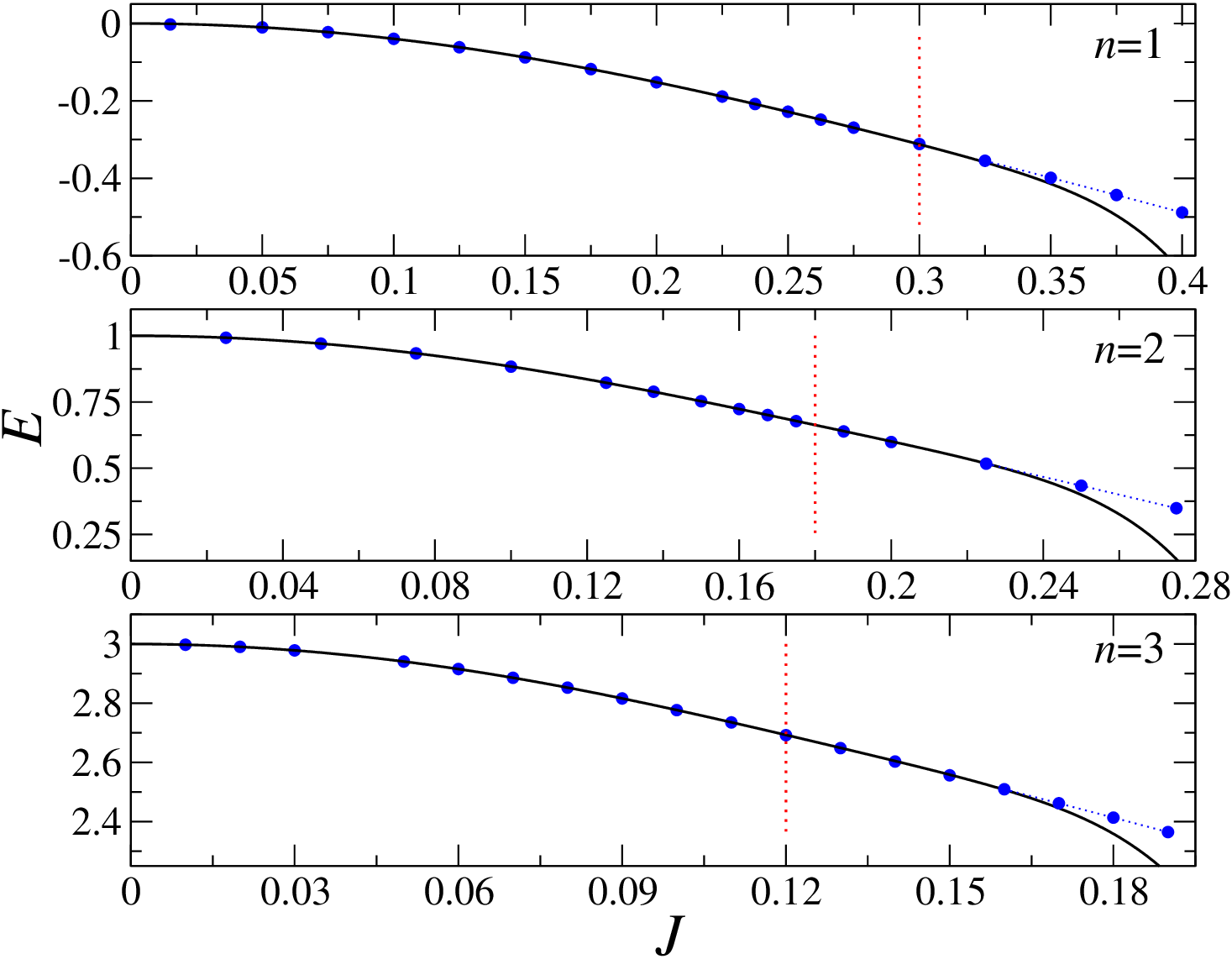}
\caption{The energy per lattice site for different filling factors. 
         Lines come from expansions (\ref{E_n=1})--(\ref{E_n=3}), while dots
	 show numerical results obtained using iTEBD code with the imaginary time evolution. 
	 Both here and in other figures we have (i) added  blue dotted lines
	 connecting the dots to facilitate 
	 quantification of the discrepancies  between perturbative expansions and numerics;  (ii)
	 drawn red vertical dotted lines at the positions of the critical
	 points; and (iii) used all the terms of the computed perturbative expansions 
	 listed in the paper to plot the perturbative results.
}
\label{fig_E}
\end{figure}

The ground state energy per lattice site $E$ 
for the unit filling factor  is
\be
\begin{aligned}
\frac{E}{4}=& - J^{2} + J^{4} +\frac{68}{9} J^{6} -\frac{1267}{81} J^{8}
+\frac{44171}{1458} J^{10} -\frac{4902596}{6561} J^{12}
-\frac{8020902135607}{2645395200} J^{14}
\\&-\frac{32507578587517774813}{466647713280000} J^{16},
\end{aligned}
\label{E_n=1}
\ee
while for $n=2$  it is given by 
\be
\begin{aligned}
\frac{E}{4}=& \frac{1}{4}-3 J^{2} +8 J^{4} +\frac{49604}{315} J^{6}
-\frac{3385322797}{13891500} J^{8} +\frac{8232891127289}{168469166250} J^{10}
\\&-\frac{7350064303936751836656911}{15282461406452625000} J^{12},
\end{aligned}
\label{E_n=2}
\ee
and finally for $n=3$ it reads
\be
\begin{aligned}
\frac{E}{4}=& \frac{3}{4}-6 J^{2} +31 J^{4} +\frac{73664}{63} J^{6}
-\frac{11207105017}{36117900} J^{8}
-\frac{76233225199535567419}{3516204203386875} J^{10}
\\&-\frac{39433892936615327274896871074109109}{1229047086250770739427475000}
J^{12}.
\end{aligned}
\label{E_n=3}
\ee
The ground state energy for an arbitrary integer filling factor was
perturbatively calculated up to the $J^4$ terms in Sec. 7.1 of Ref.
\cite{KrutitskyReview}. Our expansions, of course, match this result.

A quick inspection of  Fig. \ref{fig_E} reveals that 
there is an excellent agreement between numerics and finite-order perturbative
expansions (\ref{E_n=1})--(\ref{E_n=3}) not only in the whole Mott insulator phase, but also on
the superfluid side near the critical point
(see Ref. \cite{KnapPRA2012} for the same observation in the
$n=1$ system). 
This is  a bit surprising for  two reasons. 

First,
it is expected that the perturbative expansions break down at the critical
point in the thermodynamically-large systems undergoing a quantum phase
transition. 
This, however, does not mean that our finite-order expansions (\ref{E_n=1})--(\ref{E_n=3})
cannot accurately approximate ground state energy per lattice site across the
critical point.

Second, we find it actually more surprising that despite the fact that our
finite-order
perturbative expansions for  both $C(1)$ and $D(0)$ depart from the numerics 
on the Mott side, their combination (\ref{ECD}) works so well across the
critical point. The two-point correlation function $C(1)$ is 
depicted in Figs. \ref{fig_C_n=1}--\ref{fig_C_n=3}, while 
$D(0)$ is  given by $var(\hat n)+n^2$, where $var(\hat n)$
is plotted in Fig. \ref{fig_var}. 
It would be good to understand whether this cancellation comes as a
coincidence due to the finite-order of our perturbative expansions
(\ref{E_n=1})--(\ref{E_n=3}).

\section{Variance of  on-site number operator}
\label{sec_variance}
\begin{figure}[t]
\includegraphics[width=0.56\textwidth,clip=true]{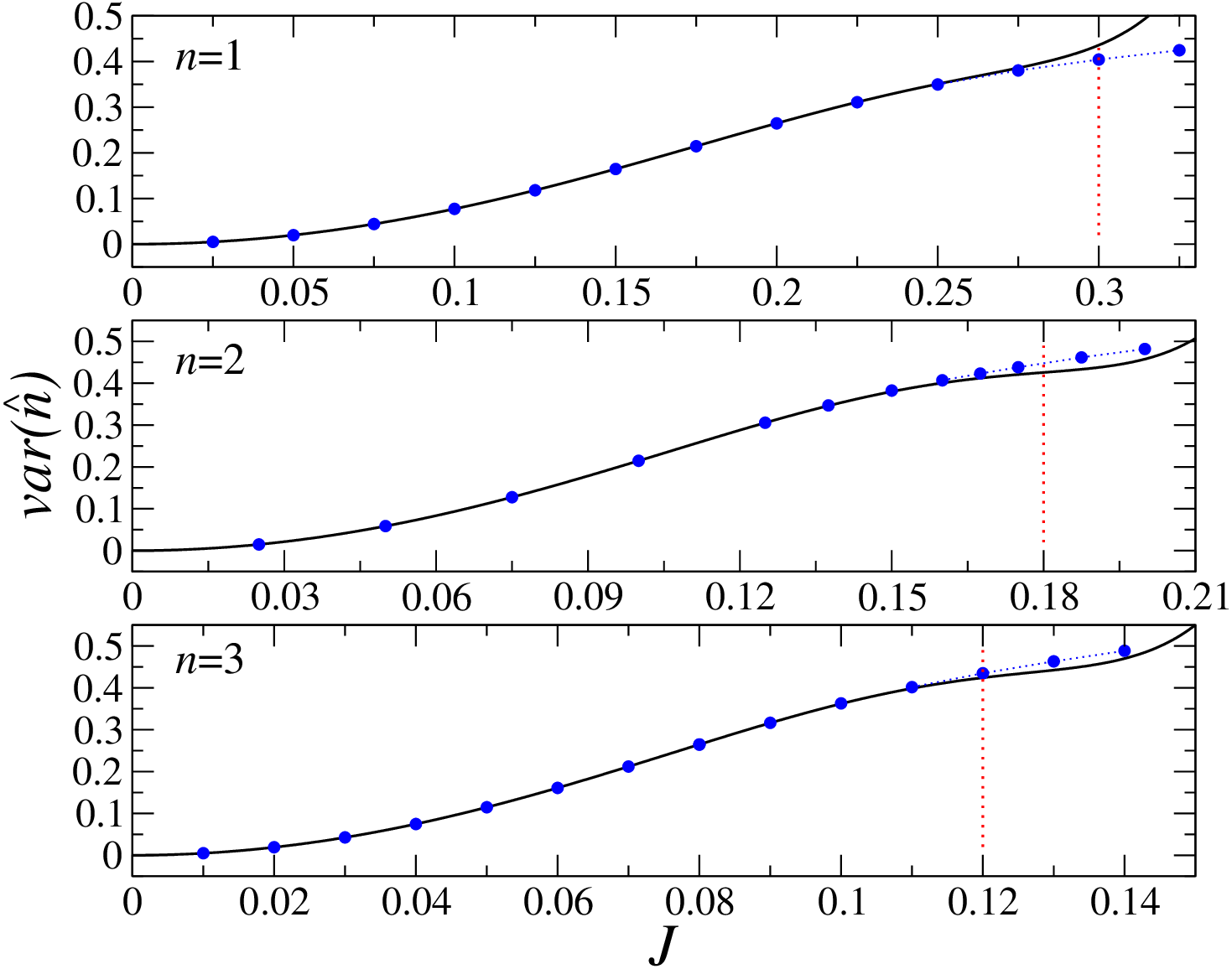}
\caption{The variance (\ref{var}) of the on-site atom number operator 
         for the filling factors  $n=1,2,3$. 
         Lines come from expansions (\ref{var_n=1})--(\ref{var_n=3}), while the dots represent numerics.
}
\label{fig_var}
\end{figure}
The most basic insight into the 
{\it local} fluctuations of the number of atoms in the ground state  is delivered by the 
variance of the on-site number operator
\be
var(\hat n)=\langle\hat n_i^2\rangle - \langle\hat
n_i\rangle^2=D(0)-n^2.
\label{var}
\ee
This quantity  is experimentally accessible due to the spectacular recent progress in 
the quantum gas microscopy  \cite{GreinerPRA2015}.

We find that for the unit filling factor 
\be
\begin{aligned}
var(\hat n)=& 8 J^{2} -24 J^{4} -\frac{2720}{9} J^{6} +\frac{70952}{81} J^{8}
-\frac{176684}{81} J^{10} +\frac{431428448}{6561} J^{12}
+\frac{104271727762891}{330674400} J^{14}
\\&+\frac{32507578587517774813}{3888730944000} J^{16},
\end{aligned}
\label{var_n=1}
\ee
for the filling factor $n=2$ 
\be
\begin{aligned}
var(\hat n)=& 24 J^{2} -192 J^{4} -\frac{396832}{63} J^{6}
+\frac{6770645594}{496125} J^{8} -\frac{32931564509156}{9359398125} J^{10}
\\&+\frac{7350064303936751836656911}{173664334164234375} J^{12}, 
\end{aligned}
\label{var_n=2}
\ee
and finally for $n=3$ 
\be
\begin{aligned}
var(\hat n)=& 48 J^{2} -744 J^{4} -\frac{2946560}{63} J^{6}
+\frac{22414210034}{1289925} J^{8}
+\frac{609865801596284539352}{390689355931875} J^{10}\\
&+\frac{39433892936615327274896871074109109}{13966444161940576584403125}
J^{12}.
\end{aligned}
\label{var_n=3}
\ee

The comparison between these perturbative expansions and numerics is
presented in Fig. \ref{fig_var}. We see there that our  expansions 
accurately match numerics in most of the  Mott phase and break
down near the critical point. It might be worth to note that these
on-site atom number fluctuations are nearly the same at the critical point
(\ref{Jc}) for the different filling factors (they equal roughly $0.4$ there).

\begin{figure}[t]
\includegraphics[width=0.56\textwidth,clip=true]{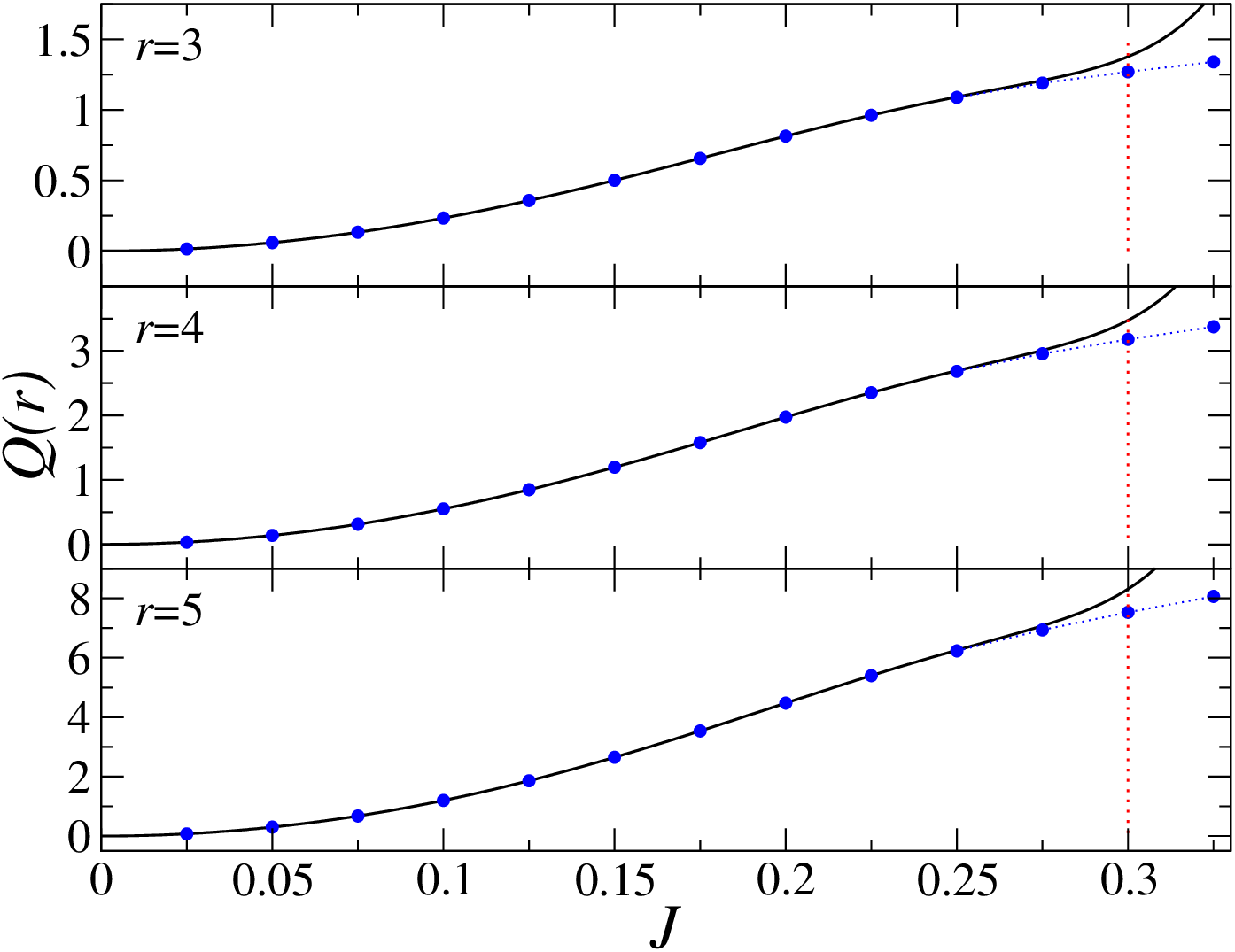}
\caption{Expectation values of the powers of the on-site number operator (\ref{Q}) for the unit filling factor.
         Lines show   expansions (\ref{Q_n=1_r=3})--(\ref{Q_n=1_r=5}), 
	 while the dots show numerics. 
}
\label{fig_Q_n=1}
\end{figure}

\section{Powers of  number operator}
\label{sec_powers}
Further characterization of the fluctuations of the occupation of individual lattice sites
comes from the study of expectation values of the integer powers of the on-site number operator
\be
Q(r)=\langle\hat n_i^r\rangle-n^r
\label{Q}
\ee
for $r>2$ (the $r=2$ case was analyzed in Sec. \ref{sec_variance}). 
Once again, we mention that these observables can be 
experimentally studied \cite{GreinerPRA2015}.

\begin{figure}[t]
\includegraphics[width=0.56\textwidth,clip=true]{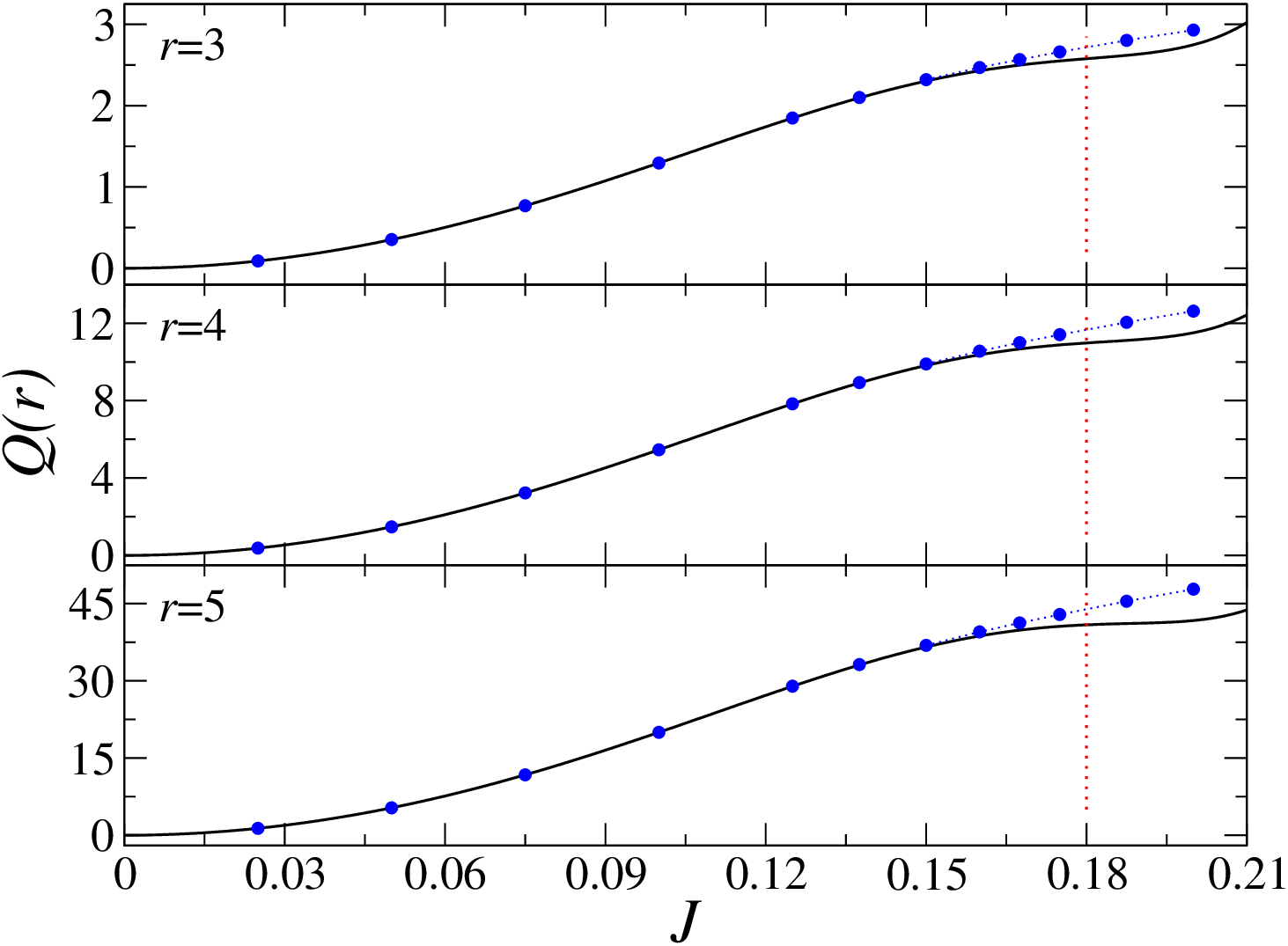}
\caption{Expectation values of the powers of the on-site number 
operator (\ref{Q}) for the $n=2$ filling factor.
         Lines show   expansions (\ref{Q_n=2_r=3})--(\ref{Q_n=2_r=5}),
	 while the dots show numerics. 
}
\label{fig_Q_n=2}
\end{figure}
For the unit filling factor, we get 
\bx
Q(3)=& 24 J^{2} -56 J^{4} -976 J^{6} +\frac{185672}{81} J^{8}
-\frac{2584369}{243} J^{10} +\frac{11909666873}{52488} J^{12}
+\frac{6518027091181469}{9258883200} J^{14}
\\&+\frac{5938172375134531873121}{181474110720000} J^{16},
\label{Q_n=1_r=3}
\ex
\bx
Q(4)=&56 J^{2} -72 J^{4} -\frac{22784}{9} J^{6} +\frac{355192}{81}
J^{8} -\frac{31533614}{729} J^{10} +\frac{16939285963}{26244} J^{12}
+\frac{488931794121599}{661348800} J^{14}
\\&+\frac{12234501340429656667403}{116661928320000} J^{16}, 
\label{Q_n=1_r=4}
\ex
\bx
Q(5)=& 120 J^{2} +40 J^{4} -\frac{18800}{3} J^{6} +\frac{601000}{81}
J^{8} -\frac{123485195}{729} J^{10} +\frac{31523026139}{17496} J^{12}
-\frac{1978940191363981}{1322697600} J^{14}
\\&+\frac{2143214705361163325357}{6666395904000} J^{16}.
\label{Q_n=1_r=5}
\ex
For two atoms per site, we obtain
\bx
Q(3)=& 144 J^{2} -1072 J^{4} -\frac{1913176}{49} J^{6}
+\frac{1770730207436}{24310125} J^{8}
-\frac{677140395560605171}{2162020966875} J^{10}
\\&+\frac{15451331550936239672643340032833}{60371453478515288484375} J^{12},
\label{Q_n=2_r=3}
\ex
\bx
Q(4)=& 600 J^{2} -3736 J^{4} -\frac{1885928848}{11025} J^{6}
+\frac{5417457952036}{40516875} J^{8}
-\frac{107844070676948560562}{32430314503125} J^{10}
\\&+\frac{59365618684278231437723679395069}{54883139525922989531250} J^{12}, 
\label{Q_n=2_r=4}
\ex
\bx
Q(5)=& 2160 J^{2} -9440 J^{4} -\frac{483627832}{735} J^{6}
-\frac{12006980573744}{24310125} J^{8}
-\frac{52636963475404293323}{2162020966875} J^{10}
\\&+\frac{480125387136897585787036245853433}{120742906957030576968750} J^{12}. 
\label{Q_n=2_r=5}
\ex
Finally, for three atoms per site we derive
\bx
Q(3)=& 432 J^{2} -6472 J^{4} -\frac{1284712}{3} J^{6}
+\frac{1195336576618}{16769025} J^{8}
+\frac{27678339796268712326815412}{3184508940200713125} J^{10}
\\&+\frac{1273450413079818438111858514006273177409357}{50089814000150161485366743625000}
J^{12}, 
\label{Q_n=3_r=3}
\ex
\bx
Q(4)=& 2640 J^{2} -36400 J^{4} -\frac{846750928}{315} J^{6}
-\frac{59064210154568}{23476635} J^{8}
-\frac{1031160890254623471701872}{974849675571646875} J^{10}
\\&+\frac{66279835521862060615675760372212019355789667}{425763419001276372625617320812500}
J^{12}, 
\label{Q_n=3_r=4}
\ex
\bx
Q(5)=& 13680 J^{2} -163280 J^{4} -\frac{101064696}{7} J^{6}
-\frac{928759047058552}{23476635} J^{8}
-\frac{4229961332321756833865450804}{9553526820602139375} J^{10}
\\&+\frac{9189183527664354899691980380144063394455799}{11353691173367369936683128555000}
J^{12}.
\label{Q_n=3_r=5}
\ex

\begin{figure}[t]
\includegraphics[width=0.56\textwidth,clip=true]{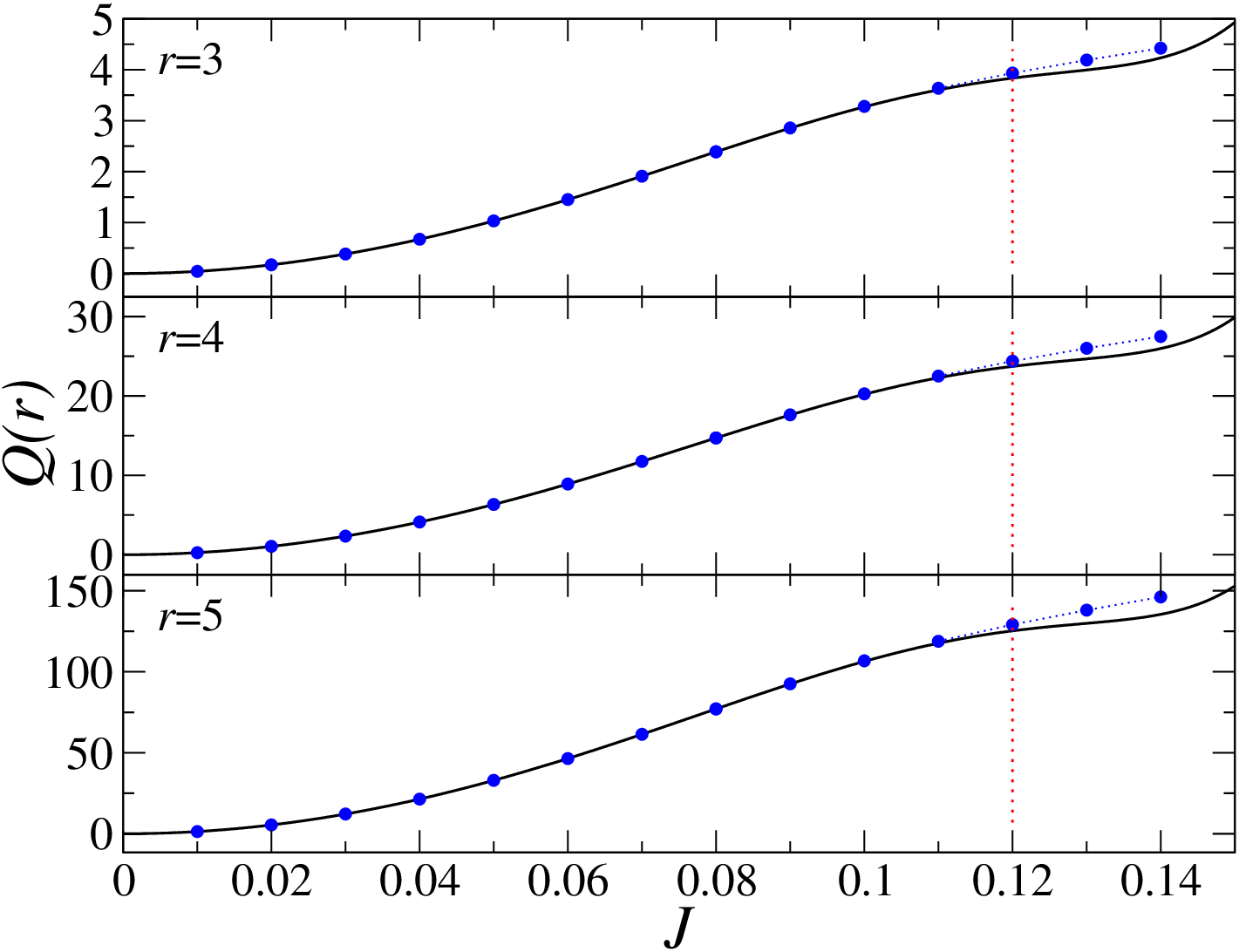}
\caption{Expectation values of the powers of 
the on-site number operator (\ref{Q}) for the  $n=3$ filling factor.
         Lines show   expansions (\ref{Q_n=3_r=3})--(\ref{Q_n=3_r=5}),
	 while the dots show numerics. 
}
\label{fig_Q_n=3}
\end{figure}

These expansions are compared to numerics in Figs.
\ref{fig_Q_n=1}--\ref{fig_Q_n=3}. They reproduce the numerics in the Mott
insulator phase in the same range of the tunneling coupling $J$ as our
expansions for the variance of the on-site number operator.

\begin{figure}[t]
\includegraphics[width=0.56\textwidth,clip=true]{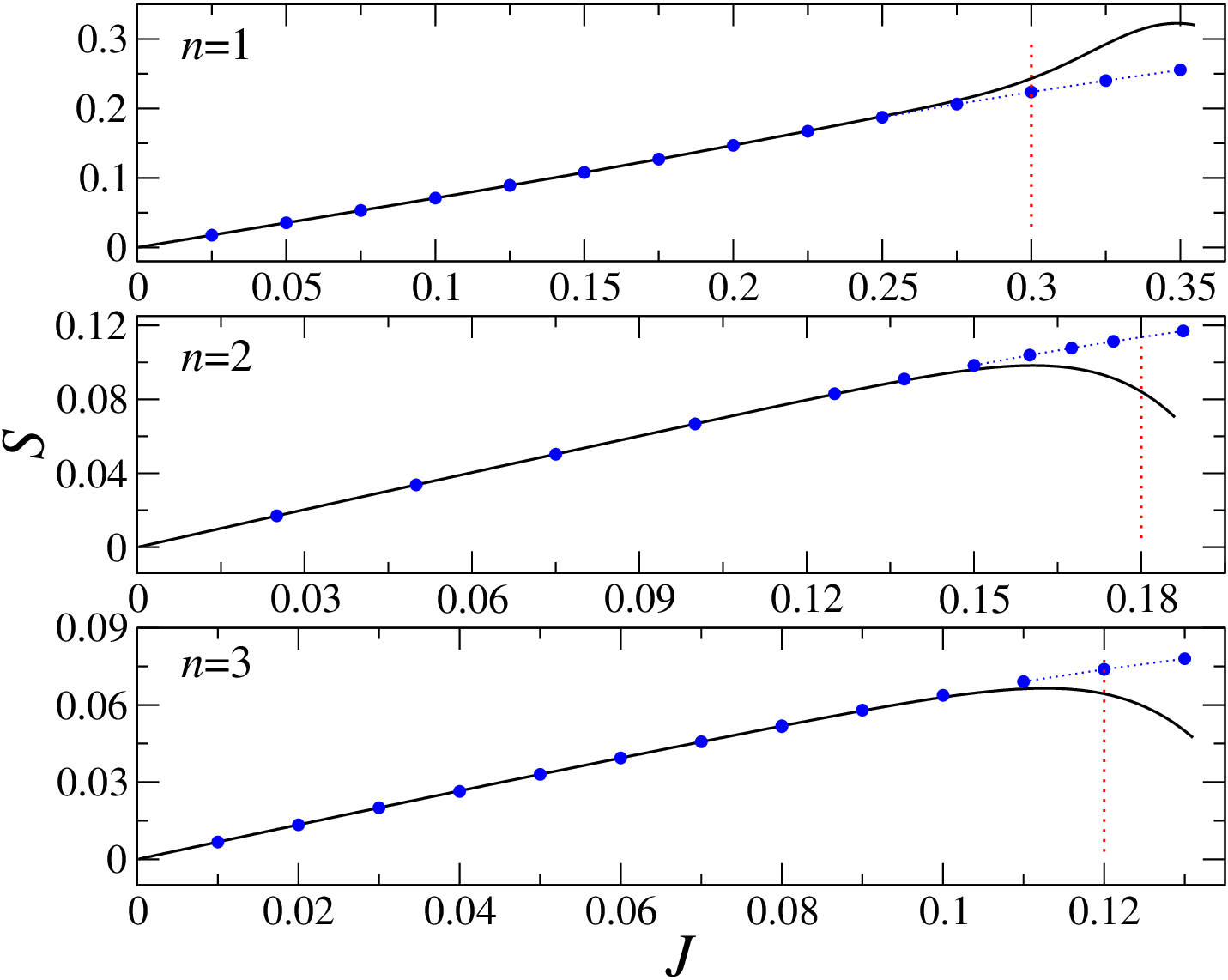}
\caption{The skewness of the on-site atom number distribution.
         Lines show  Eq. (\ref{Skew}) computed with  expansions from Secs.
	 \ref{sec_variance} and \ref{sec_powers}. Dots show numerics. 
}
\label{fig_Skew}
\end{figure}

Using expansions (\ref{Q_n=1_r=3})--(\ref{Q_n=3_r=5}) 
one can easily go further, i.e., beyond the variance,  in characterization of 
the on-site atom number distribution. For example, one can easily compute the skewness \cite{NIST,Bulmer}
\be
S= \dfrac{\la (\hat n_i-n)^3 \ra}{\la (\hat n_i-n)^2 \ra^{3/2}}
\label{Skew}
\ee
and the kurtosis \cite{NIST,Bulmer} (also referred to as excess kurtosis)
\be
K=\dfrac{\la (\hat n_i-n)^4\ra}{\la (\hat n_i-n)^2\ra^2}-3.
\label{Kurto}
\ee
The skewness is a measure of a symmetry of the distribution. It is zero for a
distribution that is symmetric around the mean. We plot  the skewness in Fig. \ref{fig_Skew}
and find it to be positive in the Mott
insulator phase, which indicates that the distribution of different numbers of
atoms is tilted towards larger-than-mean on-site occupation numbers. This is
a  somewhat expected result given the fact that the possible atom occupation numbers 
are bounded from below by zero and unbounded from above. 
Given the fact that $|S|<1/2$ in Fig.
\ref{fig_Skew},  one may conclude that the on-site  atom number distribution 
is ``fairly symmetric'' in the Mott phase according to the criteria from Ref. \cite{Bulmer}.

\begin{figure}[t]
\includegraphics[width=0.56\textwidth,clip=true]{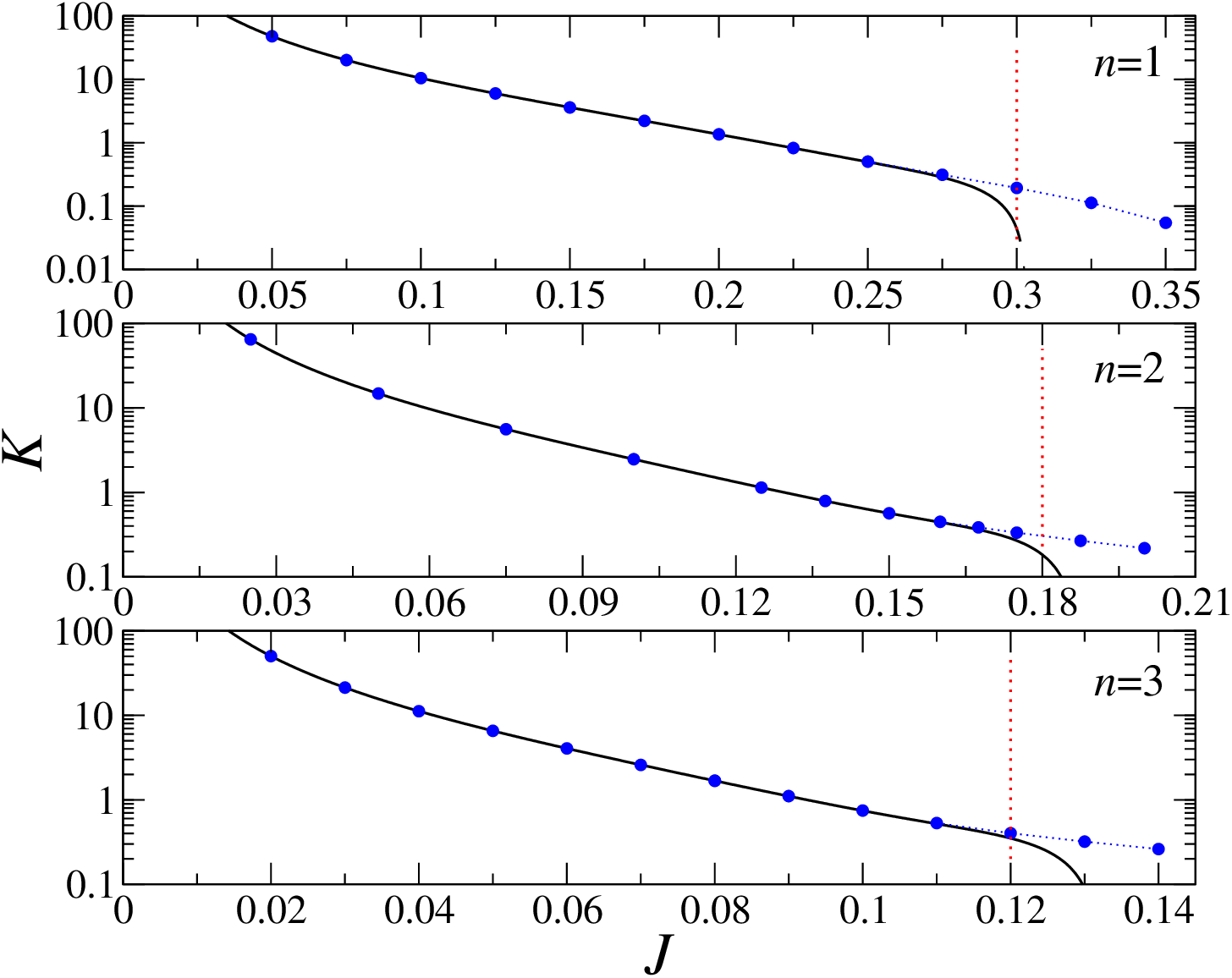}
\caption{The kurtosis of the on-site atom number distribution.
         Lines show  Eq. (\ref{Kurto}) computed with  expansions from Secs.
	 \ref{sec_variance} and \ref{sec_powers}. Dots show numerics. 
	 }
\label{fig_Kurto}
\end{figure}

The kurtosis quantifies whether the distribution is peaked or flat relative to the 
normal (Gaussian) distribution. It is calibrated such that it equals zero for the  normal
distribution of arbitrary mean and variance. $K>0$ ($K<0$) indicates that the studied distribution 
is peaked (flattened) relative to the normal distribution.
We plot the kurtosis in Fig. \ref{fig_Kurto}. As $J\to0$ one easily finds from our
expansions that $K\sim J^{-2}$. This singularity reflects the strong 
suppression of the local atom number fluctuations in the deep Mott insulator limit. 
The kurtosis monotonically decays in the Mott phase (Fig. \ref{fig_Kurto}).

To put these results in context, we compare them to the on-site atom number
distribution in the deep superfluid limit of $J\to\infty$ (the Poisson distribution \cite{SvistunovPRA2007}). 
The probability of
finding $s$ atoms in a lattice site is then given in the thermodynamic limit by $\exp(-n) n^s/s!$, where 
$n$ is the mean occupation.
One then  finds that $S=1/\sqrt{n}$ and 
$K=1/n$ for the Poisson distribution. 
Keeping in mind that the Gaussian distribution is characterized by $S=K=0$, we 
can try to see  whether the on-site atom
number distribution near the critical point is  Gausssian-like
or Poissonian-like.

We see from Figs. \ref{fig_Skew} and \ref{fig_Kurto}  that at the critical
point (\ref{Jc}) we have  $S\approx0.22, 0.11, 0.07$ and $K\approx0.19,0.3,0.4$ for $n=1,2,3$,
respectively. Therefore, the real distribution lies somehow between Poissonian and Gaussian. The
skewness suggests that for these filling factors the
distribution at the critical point is more Gaussian than Poissonian. 
On the other hand, the kurtosis for $n=1$ ($n=2,3$) is more Gaussian
(Poissonian). From this we conclude that for the unit filling factor 
the on-site atom number distribution at the critical point 
is better approximated by the Gaussian distribution.

\section{Two-point correlations}
\label{sec_twopoint}
The two-point correlation functions play a special role in the  cold atom
realizations of the Bose-Hubbard model \cite{BlochNature2002,ProkofevPRA2002,ZwergerReview2008}. 
Their Fourier transform
provides the quasi-momentum distribution of a cold atom cloud, which is  visible 
through the time-of-flight images that are taken after releasing the cloud from
the trap. 

\begin{figure}[t]
\includegraphics[width=0.56\textwidth,clip=true]{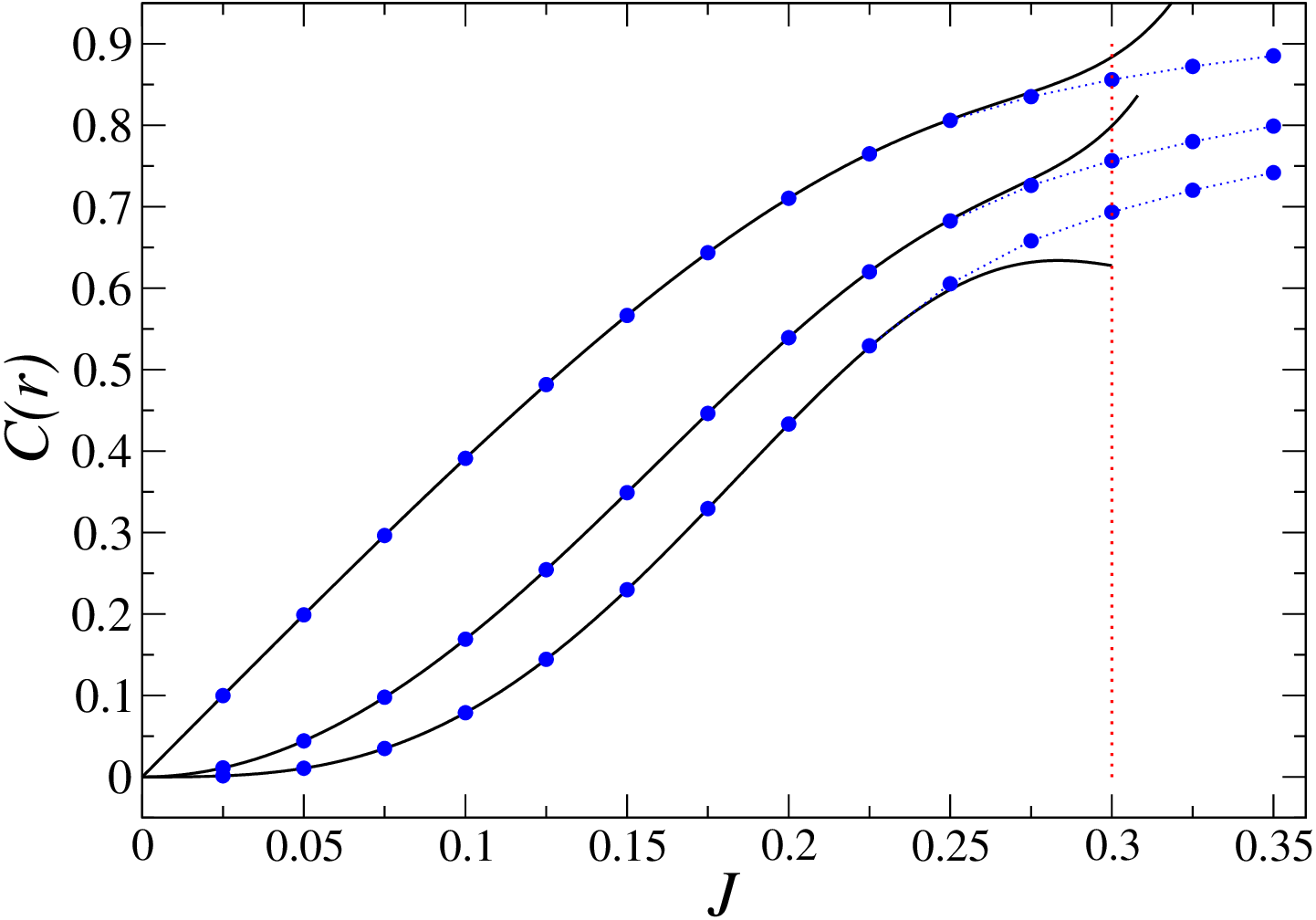}
\caption{
The two-point correlation functions for the unit filling factor.
         Lines  from top to bottom correspond to $r=1,2,3$, respectively.
	 They depict perturbative   expansions (\ref{C_n=1_r=1})--(\ref{C_n=1_r=3}). 
	 The numerics is presented with dots.
}
\label{fig_C_n=1}
\end{figure}

\begin{figure}[t]
\includegraphics[width=0.56\textwidth,clip=true]{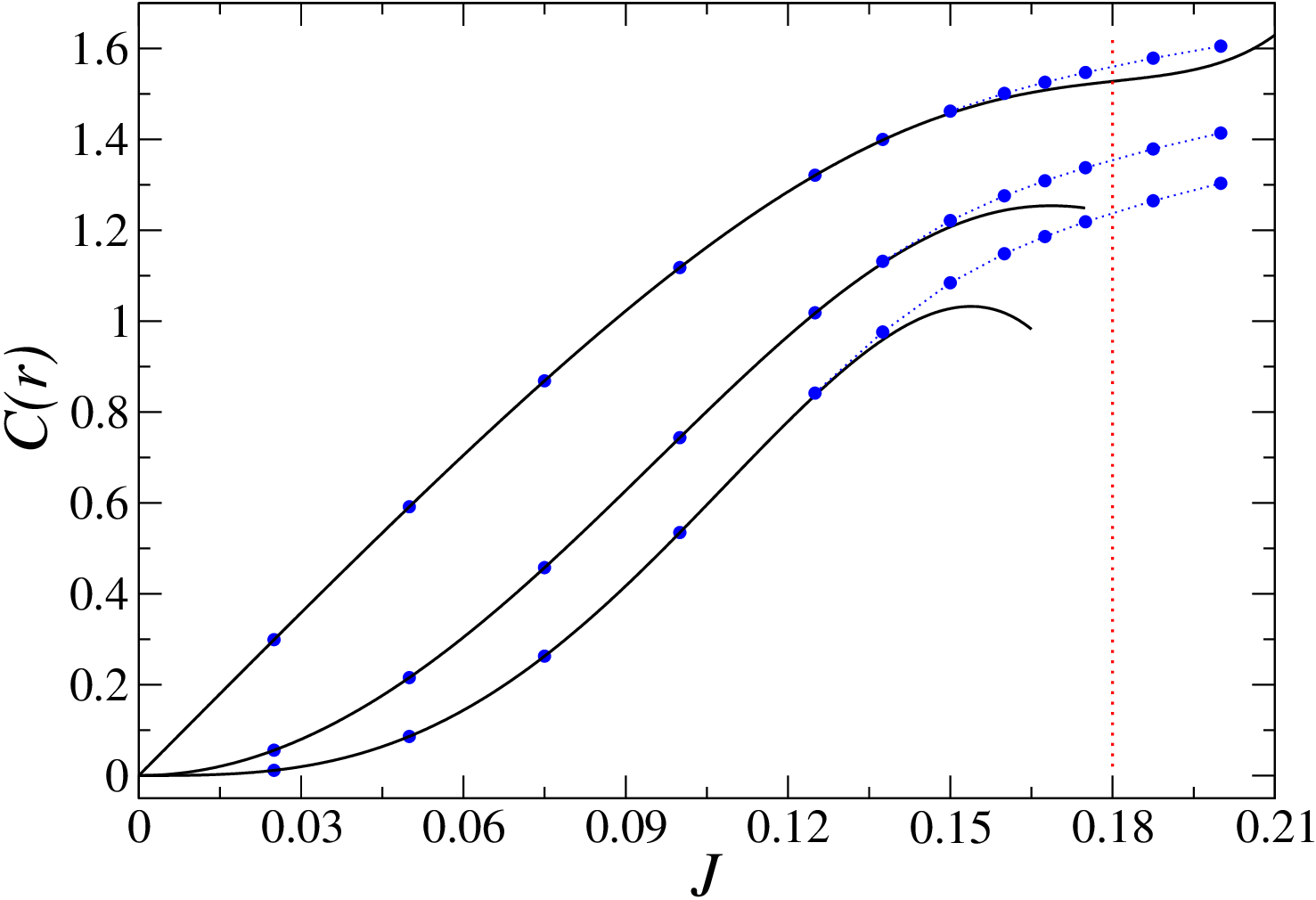}
\caption{
The two-point correlation functions for the $n=2$ filling factor.
         Lines  from top to bottom correspond to $r=1,2,3$, respectively.
They depict perturbative   expansions  (\ref{C_n=2_r=1})--(\ref{C_n=2_r=3}). 
	 The numerics is presented with dots.
}
\label{fig_C_n=2}
\end{figure}

\begin{figure}[t]
\includegraphics[width=0.56\textwidth,clip=true]{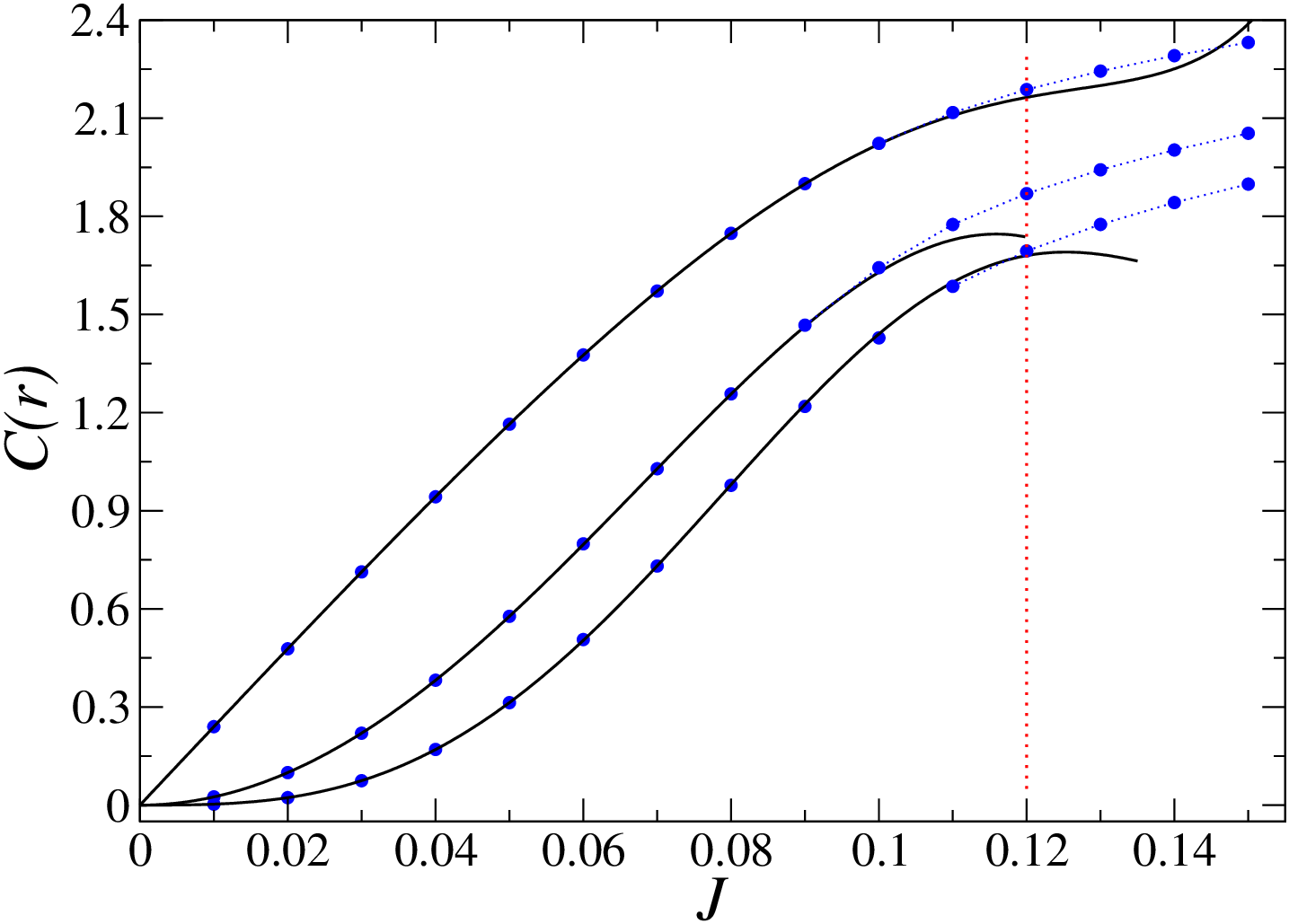}
\caption{The two-point correlation functions for the $n=3$ filling factor.
         Lines  from top to bottom correspond to $r=1,2,3$, respectively.
They depict perturbative   expansions  (\ref{C_n=3_r=1})--(\ref{C_n=3_r=3}). 
	 The numerics is presented with dots.
}
\label{fig_C_n=3}
\end{figure}

For the filling factor $n=1$, they are given by 
\bx
C(1)=& 4 J -8 J^{3} -\frac{272}{3} J^{5} +\frac{20272}{81} J^{7}
-\frac{441710}{729} J^{9} +\frac{39220768}{2187} J^{11}
+\frac{8020902135607}{94478400} J^{13}\\&+\frac{32507578587517774813}{14582741040000} J^{15},
\label{C_n=1_r=1}
\ex
\bx
C(2)=& 18 J^{2} -\frac{320}{3} J^{4} -\frac{1826}{9} J^{6} +\frac{234862}{243}
J^{8} +\frac{345809}{2916} J^{10} +\frac{4434868108963}{220449600} J^{12}
\\&+\frac{94620702880069301837}{38887309440000} J^{14},
\label{C_n=1_r=2}
\ex
\bx
C(3)=& 88 J^{3} -\frac{8324}{9} J^{5} +\frac{126040}{81} J^{7}
+\frac{7883333}{486} J^{9} -\frac{220980576341}{1049760} J^{11}
\\&+\frac{82283484127688477}{61725888000} J^{13},
\label{C_n=1_r=3}
\ex
and for the filling factor $n=2$ they are 
\bx
C(1)=& 12 J -64 J^{3} -\frac{198416}{105} J^{5}
+\frac{13541291188}{3472875} J^{7} -\frac{16465782254578}{16846916625} J^{9}
\\&+\frac{7350064303936751836656911}{636769225268859375} J^{11},
\label{C_n=2_r=1}
\ex
\bx
C(2)=& 90 J^{2} -\frac{4520}{3} J^{4} -\frac{12971657}{2205} J^{6}
-\frac{290416211186}{16372125} J^{8}
+\frac{15957686927590379575531}{10511745940946250} J^{10}
\\&+\frac{651222142925783091305873230764520129}{10926355845430110013929000000}
J^{12},
\label{C_n=2_r=2}
\ex
\bx
C(3)=& 744 J^{3} -\frac{201172}{9} J^{5} +\frac{467115289252}{3472875} J^{7}
+\frac{934116436332193243}{617720276250} J^{9}
\\&-\frac{165376430398934085307383814830617}{1931886511312489231500000} J^{11},
\label{C_n=2_r=3}
\ex
and finally for $n=3$ they read
\bx
C(1)=& 24 J -248 J^{3} -\frac{294656}{21} J^{5}
+\frac{44828420068}{9029475} J^{7}
+\frac{304932900798142269676}{703240840677375} J^{9}\\
&+\frac{39433892936615327274896871074109109}{51210295260448780809478125} J^{11},
\label{C_n=3_r=1}
\ex
\bx
C(2)=& 252 J^{2} -\frac{24920}{3} J^{4} -\frac{2559347}{45} J^{6}
-\frac{912812009912144}{774728955} J^{8}
+\frac{728914146234298491592146132346}{8932547577263000315625} J^{10},
\label{C_n=3_r=2}
\ex
\bx
C(3)=& 2928 J^{3} -\frac{1563584}{9} J^{5} +\frac{60570509140}{27783} J^{7}
+\frac{21318637245947810350682}{678565723460625} J^{9}.
\label{C_n=3_r=3}
\ex
Expansions up to the order $J^3$ for $C(1)$, $C(2)$, and $C(3)$ at arbitrary
integer filling factors are listed in Sec. 7.1 of Ref.
\cite{KrutitskyReview} and agree with our results.

We see in Figs. \ref{fig_C_n=1}--\ref{fig_C_n=3} that the above
perturbative expansions break down within the Mott insulator phase
(the larger $r$, the deeper in the Mott  phase the expansion breaks down).
We notice that  it is instructive 
to compare the value of the
correlations $C(r)$ around the critical point to their deep superfluid limit. 
$C(r)$ in the $J\to\infty$ limit tends to $n$ (see e.g. Appendix B of Ref.
\cite{BDJZPRA2006}). Therefore, the three correlation functions $C(r=1,2,3)$ 
reach   at least $50\%$ of their deep superfluid value near the
critical point, which well illustrates the significance  of  quantum
fluctuations at the critical point.

The ground state quasi-momentum distribution is defined as
\bee
\tilde n(k) = \frac{1}{M}\sum_{m,s=1}^M\la \hat a^\dag_m\hat a_s\ra \exp\BB{ik(m-s)},
\eee
where $M$ stands for the number of lattice sites (we skip the prefactor proportional to the 
squared modulus of the Fourier transform of the Wannier functions; see Ref. \cite{ProkofevPRA2002}
for  details). Taking the limit of
$M\to\infty$ at the fixed integer filling factor $n$, one  gets 
\bee
\tilde n(k)=n + 2\sum_{r=1}^\infty C(r)\cos(rk).
\eee

Using Eqs. (\ref{C_n=1_r=1})--(\ref{C_n=1_r=3}) and (\ref{C_n=1_r=4})--(\ref{C_n=1_r=8}) for $n=1$, 
      Eqs. (\ref{C_n=2_r=1})--(\ref{C_n=2_r=3}) and (\ref{C_n=2_r=4})--(\ref{C_n=2_r=7}) for $n=2$,
      and 
      Eqs. (\ref{C_n=3_r=1})--(\ref{C_n=3_r=3}) and (\ref{C_n=3_r=4})--(\ref{C_n=3_r=6}) for $n=3$
the state-of-the-art 
high-order perturbative quasi-momentum distributions for different filling factors can be obtained. These results 
can be compared to Ref. \cite{TrivediPRA2009}, where an expression with terms
up to $J^3$ for an arbitrary filling factor is computed. As expected, we find
these results in agreement with our findings.

\section{Density-density correlations}
Similarly as the observables from Secs. \ref{sec_variance} and \ref{sec_powers}, the
density-density correlations can be experimentally approached 
through the technique discussed in Ref. \cite{GreinerPRA2015}.
\label{sec_density}
\begin{figure}[t]
\includegraphics[width=0.56\textwidth,clip=true]{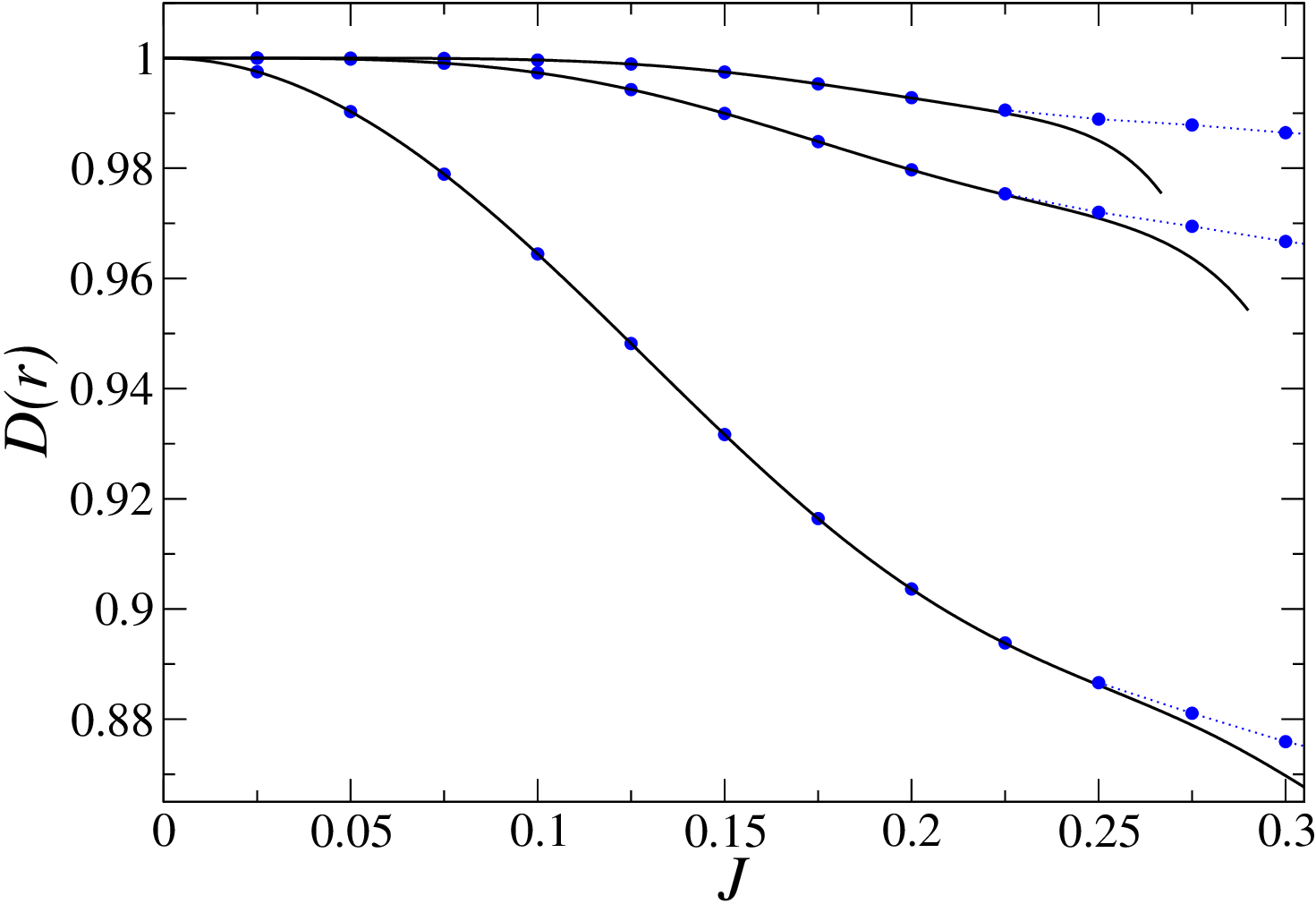}
\caption{
The density-density correlation functions for the unit filling factor.
         Lines from bottom to top illustrate perturbative results for
	 $r=1,2,3$, respectively. Dots show numerics. Perturbative
	 expansions are given by Eqs. (\ref{D_n=1_r=1})--(\ref{D_n=1_r=3}). 
}
\label{fig_D_n=1}
\end{figure}

\begin{figure}[t]
\includegraphics[width=0.56\textwidth,clip=true]{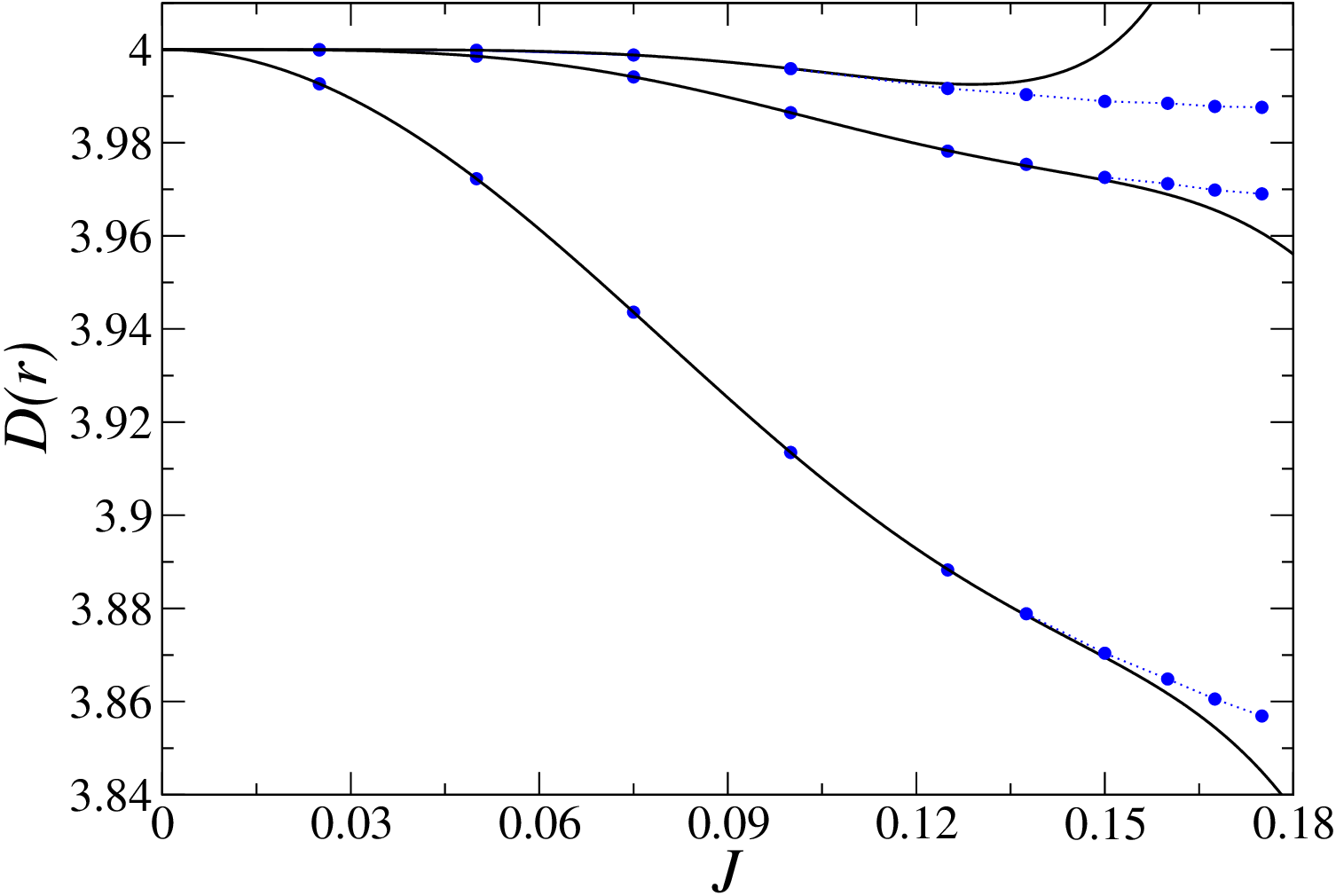}
\caption{The density-density correlation functions for the $n=2$ filling factor.
         Lines from bottom to top illustrate perturbative results for
	 $r=1,2,3$, respectively. Dots show numerics. Perturbative
	 expansions are given by Eqs. (\ref{D_n=2_r=1})--(\ref{D_n=2_r=3}). 
}
\label{fig_D_n=2}
\end{figure}

\begin{figure}[t]
\includegraphics[width=0.56\textwidth,clip=true]{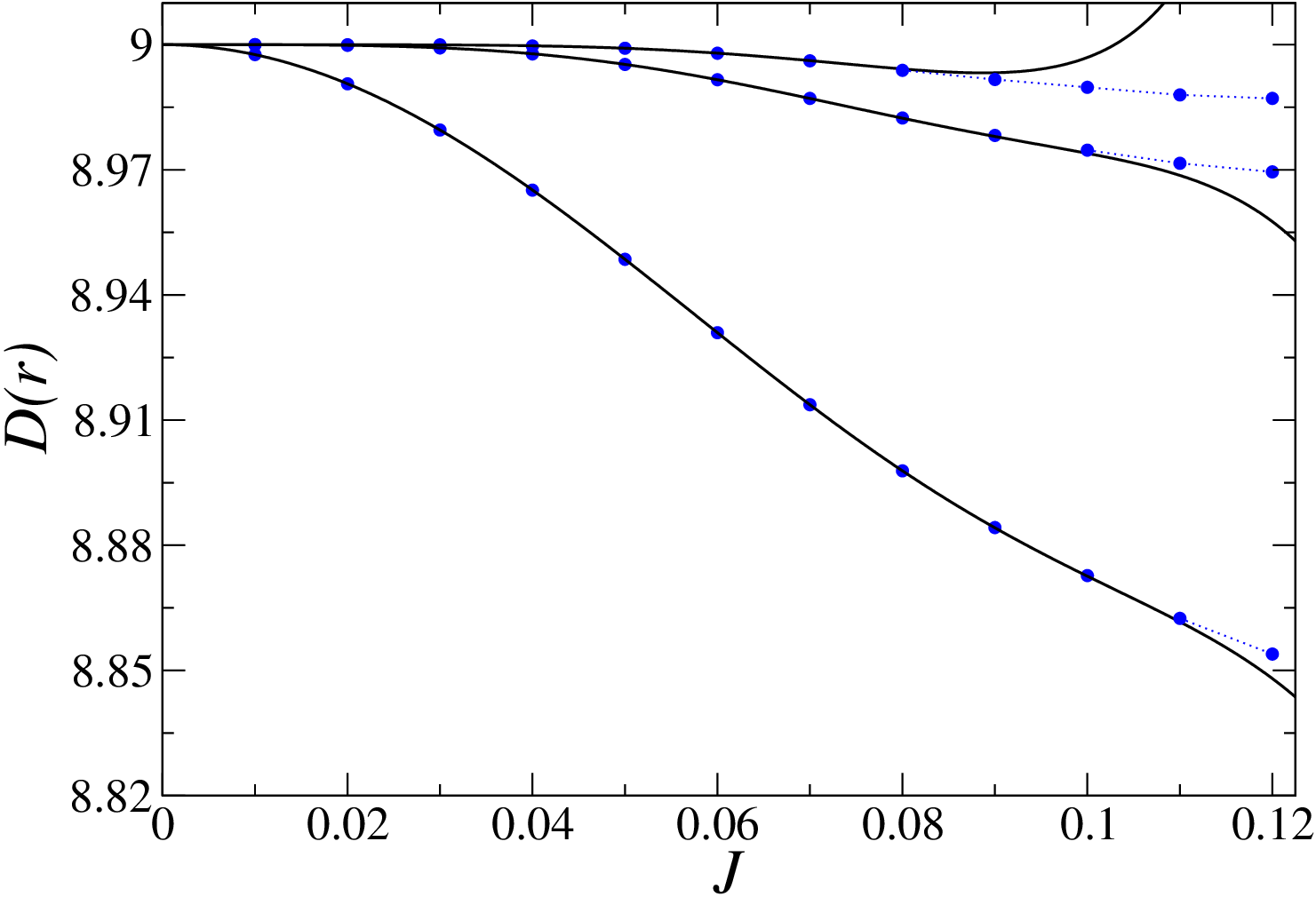}
\caption{
The density-density correlation functions for the $n=3$ filling factor.
         Lines from bottom to top illustrate perturbative results for
	 $r=1,2,3$, respectively. Dots show numerics. Perturbative
	 expansions are given by Eqs. (\ref{D_n=3_r=1})--(\ref{D_n=3_r=3}). 
}
\label{fig_D_n=3}
\end{figure}

The density-density correlations are given for $n=1$ by
\bx
D(1)=& 1-4 J^{2} +\frac{136}{3} J^{4} -\frac{2008}{27} J^{6} -\frac{150638}{81}
J^{8} +\frac{4897282}{729} J^{10} -\frac{415922848153}{14696640} J^{12}
\\&+\frac{1022120948444278027}{7777461888000} J^{14}
+\frac{4588274318283441920855291}{2515231174579200000} J^{16}, 
\label{D_n=1_r=1}
\ex
\bx
D(2)=& 1-\frac{100}{3} J^{4} +\frac{2128}{3} J^{6} -\frac{1156462}{243} J^{8}
-\frac{6848011}{729} J^{10} +\frac{10808763042127}{44089920} J^{12}
\\&-\frac{5150051155340205251}{3888730944000} J^{14}
-\frac{10173100607048978123860781}{15091387047475200000} J^{16}, 
\label{D_n=1_r=2}
\ex
\bx
D(3)=& 1-\frac{13064}{27} J^{6} +\frac{3727066}{243} J^{8}
-\frac{1588041877}{8748} J^{10} +\frac{1710030328933}{2755620} J^{12}
\\&+\frac{2208787916976404357}{370355328000} J^{14}
-\frac{73297040097456632572895911}{1006092469831680000} J^{16},
\label{D_n=1_r=3}
\ex
for $n=2$ they read 
\bx
D(1)=& 4-12 J^{2} +\frac{1100}{3} J^{4} -\frac{80553632}{33075} J^{6}
-\frac{8915569805768}{121550625} J^{8}
-\frac{185683648947492811}{6486062900625} J^{10}
\\&+\frac{331686439652436848222471319678887}{72445744174218346181250000} J^{12},
\label{D_n=2_r=1}
\ex
\bx
D(2)=& 4-\frac{812}{3} J^{4} +\frac{189419192}{11025} J^{6}
-\frac{140772979852859}{364651875} J^{8}
+\frac{569733162420769673609}{324303145031250} J^{10}
\\&+\frac{8056033392249986400009146407648}{503095445654294070703125} J^{12}, 
\label{D_n=2_r=2}
\ex
\bx
D(3)=& 4-\frac{7827256}{675} J^{6} +\frac{79648906064158}{72930375} J^{8}
-\frac{79481781249654943885127}{1945818870187500} J^{10}
\\&+\frac{1369141430325035192671991031520835051}{2028480836878113693075000000}
J^{12}.
\label{D_n=2_r=3}
\ex
and finally for $n=3$ they can be written as
\bx
D(1)=& 9 -24 J^{2} +\frac{4276}{3} J^{4} -\frac{134562824}{6615} J^{6}
-\frac{3678796866393562}{4108411125} J^{8}
-\frac{2257554577848066943151996417}{200624063232644926875} J^{10}\\
&+\frac{1542480719505910230376228320731168033769541}{3193225642509572794692129906093750}
J^{12},
\label{D_n=3_r=1}
\ex
\bx
D(2)=& 9 -\frac{3160}{3} J^{4} +\frac{294843464}{2205} J^{6}
-\frac{76035818562996449}{12325233375} J^{8}
+\frac{83670951564711862884744588592}{1003120316163224634375} J^{10}\\
&-\frac{143706393091669463828236051561683582721397}{132705481247151077182010593500000}
J^{12},
\label{D_n=3_r=2}
\ex
\bx
D(3)=& 9 -\frac{12148432}{135} J^{6} +\frac{246576902129764}{14586075}
J^{8} -\frac{1185488040768577918685665169}{926242212523753125} J^{10}\\
&+\frac{13080624640958701853202057691706510935349239}{285200376364491350084145573750000}
J^{12}.
\label{D_n=3_r=3}
\ex
The correlation functions $D(1)$ and $D(2)$ were computed for an arbitrary
integer filling factor up to the order $J^4$ in Sec. 7.1 of Ref.
\cite{KrutitskyReview}. These results agree with our expansions.

The comparison between our  perturbative expansions and numerics is presented
in Figs. \ref{fig_D_n=1}--\ref{fig_D_n=3} for different filling factors.
The expansions break down near the critical point on the Mott side of the
transition. Comparing Figs. \ref{fig_C_n=1}--\ref{fig_C_n=3} to Figs.
\ref{fig_D_n=1}--\ref{fig_D_n=3}, we see that expansions for the two-point and
 density-density correlations break down in similar  distances from the critical
point. Moreover, this comparison shows that the two-point correlations 
change more appreciably within the Mott phase than the density-density
correlations. We attribute it to the constraints that are imposed on the
density-density correlations due to the atom number conservation.

\section{Summary}
\label{sec_sum}
We have computed  state-of-the-art high-order perturbative expansions for several observables
characterizing ground state properties of the one-dimensional Bose-Hubbard model in the Mott phase. 
As compared to our former results for the filling factor $n=1$  \cite{BDJZPRA2006}, we have extended our analysis
by considering the filling factors $n=2$ and $3$ (we have also  enlarged the number of terms for the $n=1$ case). 
We have characterized the on-site atom number distribution  by giving the predictions for the skewness and  kurtosis. 
Those may serve as useful benchmarks for experimental in-situ observations \cite{GreinerPRA2015}. 
We have also derived  in a simple way an important sum rule applicable to both equilibrium and non-equilibrium density-density correlations. 
That sum rule  allows for verification of our perturbative expansions and it  may be useful for checking the consistency of experimental data.
We have also carefully established the range of applicability of our perturbative expansions through numerical simulations. 
The expansions discussed in this work can be  easily typed or imported
into computer software such as 
{\fontfamily{cmtt}\selectfont Mathematica}
or 
{\fontfamily{cmtt}\selectfont Maple}
and used for benchmarking approximate approaches, comparing   theoretical 
predictions to experimental measurements, testing Pad\'e approximations, etc.

\section{Acknowledgment}
BD thanks Eddy  Timmermans for a discussion about sum rules that  happened
about a decade ago. JZ acknowledges the collaboration with Dominique Delande on developing the implementation of the iTEBD code. We acknowledge  support of Polish National Science Centre via projects
DEC-2013/09/B/ST3/00239 (BD) and DEC-2012/04/A/ST2/00088 (JZ).
Support from the EU Horizon 2020-FET QUIC 641122 is also acknowledged (JZ). 

\appendix
\section{iTEBD simulations}
\label{sec_appitebd}
There are two factors that have to be taken care of to assure the convergence of results 
in the  numerical implementation of iTEBD. The first one is the
maximal allowed number of bosons per site assumed in the variational ansatz,
$N_{max}$. We take $ N_{max}=6$ for the filling factor $n=1$ up to
$N_{max}=12$ for $n=3$. We have checked that these values lead to
converged results. The second important factor is the number of Schmidt
decomposition eigenvalues, $\chi$, kept during each step of the procedure
\cite{Vidal07,Schollwoeck11}. $\chi$ may be quite small deep in the Mott regime
(of about 20) while it must be significantly increased close to the critical
point and in the superfluid regime. We have found that for reliable
energy, particle number variance, as well as
two-point correlations with small $r$ the choice of $\chi=150$ was largely enough (with
the relative error of the order of $10^{-7}$ in energy and   $10^{-5}$ in particle
number variance). Let us note that the numerical studies of long-range correlations
($r$ of the order of a hundred) require taking $\chi>r$ at least \cite{KubaDom}.

\section{One atom per site}
\label{sec_appone}
Our remaining perturbative expansions for the $n=1$ filling factor  are listed below.

The two-point correlations:
\bx
C(4)= 450 J^{4} -\frac{186608}{27} J^{6} +\frac{7565704}{243} J^{8}
+\frac{1493509507}{17496} J^{10} -\frac{858313783040137}{440899200} J^{12},
\label{C_n=1_r=4}
\ex
\bx
C(5)= 2364 J^{5} -\frac{3894512}{81} J^{7} +\frac{250517014}{729} J^{9}
-\frac{25842700043}{209952} J^{11},
\label{C_n=1_r=5}
\ex
\bx
C(6)= 12642 J^{6} -\frac{78008768}{243} J^{8} +\frac{6836492080}{2187} J^{10},
\label{C_n=1_r=6}
\ex
\bx
C(7)= 68464 J^{7} -\frac{1522020908}{729} J^{9},
\label{C_n=1_r=7}
\ex
\bx
C(8)= 374274 J^{8}.
\label{C_n=1_r=8}
\ex

The density-density correlations:
\bx
D(4)=& 1-\frac{741706}{81} J^{8} +\frac{93328235}{243} J^{10}
-\frac{288653212433561}{44089920} J^{12}
+\frac{32675495835088308133}{648121824000} J^{14} \\
&-\frac{136868524145553747592735387}{5030462349158400000} J^{16}, 
\label{D_n=1_r=4}
\ex
\bx
D(5)=& 1-\frac{1738824899}{8748} J^{10} +\frac{91423339623697}{8817984} J^{12}
-\frac{897923823504590743589}{3888730944000} J^{14}\\&
+\frac{40404128939318210395355039327}{15091387047475200000} J^{16}, 
\label{D_n=1_r=5}
\ex
\bx
D(6)=& 1-\frac{369347437555}{78732} J^{12}
+\frac{22768945554355275259}{77774618880} J^{14}
-\frac{20246612891148030348297322711}{2515231174579200000} J^{16}, 
\label{D_n=1_r=6}
\ex
\bx
D(7)=& 1-\frac{1771595060952703}{15116544} J^{14}
+\frac{4869453765809764188858679}{571643448768000} J^{16},
\label{D_n=1_r=7}
\ex
\bx
D(8)=& 1-\frac{415126490285461535}{136048896} J^{16}. 
\label{D_n=1_r=8}
\ex

\section{Two atoms per site}
\label{sec_apptwo}
Our remaining perturbative expansions for the $n=2$ filling factor  are listed below.

The two-point correlations:
\bx
\label{C_n=2_r=4}
C(4)= 6450 J^{4} -\frac{7683200}{27} J^{6} +\frac{291093977979158}{72930375}
J^{8} +\frac{10163319115920107956583}{1712320605765000} J^{10},
\ex
\bx
\label{C_n=2_r=5}
C(5)= 57492 J^{5} -\frac{272806072}{81} J^{7} +\frac{2786470218115003978}{38288446875} J^{9},
\ex
\bx
\label{C_n=2_r=6}
C(6)= 521850 J^{6} -\frac{9291088760}{243} J^{8},
\ex
\bx
\label{C_n=2_r=7}
C(7)= 4797840 J^{7}.
\ex

The density-density correlations:
\bx
D(4)=& 4-\frac{1586483355826}{2480625} J^{8}
+\frac{25731787904762281349459}{324303145031250} J^{10}
-\frac{4867179143263377024841830332580901}{1169827472248047112500000} J^{12}, 
\label{D_n=2_r=4}
\ex
\bx
D(5)=& 4-\frac{13200913614880820989}{328186687500} J^{10}
+\frac{43439587507699792960803904702410359}{7018964833488282675000000} J^{12}, 
\label{D_n=2_r=5}
\ex
\bx
D(6)=& 4-\frac{231224301660078686005531}{84234583125000} J^{12}.
\label{D_n=2_r=6}
\ex

\section{Three atoms per site}
\label{sec_appthree}
Our remaining perturbative expansions for the $n=3$ filling factor  are listed below.

The two-point correlations:
\bx
\label{C_n=3_r=4}
C(4)= 35700 J^{4} -\frac{84091952}{27} J^{6}
+\frac{919114524688124}{10418625} J^{8},
\ex
\bx
\label{C_n=3_r=5}
C(5)= 447624 J^{5} -\frac{4203103112}{81} J^{7},
\ex
\bx
\label{C_n=3_r=6}
C(6)= 5715948 J^{6}.
\ex

The density-density correlations:
\bx
D(4)=& 9 -\frac{977276150432}{99225} J^{8}
+\frac{21533082002709807426464}{8844631228125} J^{10} \\
&-\frac{13390761501371812933197864158162939174976229}{52141897063868224669123567500000}
J^{12},
\label{D_n=3_r=4}
\ex
\bx
D(5)=& 9 -\frac{20144790858435740956}{16409334375} J^{10}
+\frac{74244526197167849189627032046547619}{197408385941857950234375}
J^{12},
\label{D_n=3_r=5}
\ex
\bx
D(6)=& 9 -\frac{242194363655594937438210358}{1459364152640625} J^{12}.
\label{D_n=3_r=6}
\ex


\end{document}